\documentclass[prd,email,preprint,showpacs,showkeys,preprintnumbers,amsmath,amssymb,nofootinbib]{revtex4}

\usepackage{amsmath}
\usepackage{latexsym}
\def\[{\left\lbrack}
\def\]{\right\rbrack}

\def\({\left(}
\def\){\right)}
\newcommand{\be}{\begin{equation}}
\newcommand{\ee}{\end{equation}}
\newcommand{\ea}{\end{eqnarray}}
\newcommand{\ba}{\begin{eqnarray}}

\newcommand{\p}{\partial}
\def\ni{\noindent}

\begin{document}

\pagestyle{myheadings}
\markright{Symmetry considerations on radiation damping}


\title{Symmetry considerations on radiation damping}


\author{Everton M. C. Abreu$^{a,b}$} 
\email{evertonabreu@ufrrj.br}
\author{Albert C. R. Mendes$^b$}
\email{albert@fisica.ufjf.br}
\author{Wilson Oliveira$^b$} 
\email{wilson@fisica.ufjf.br}

\affiliation{${}^{a}$Grupo de F\' isica Te\'orica e Matem\'atica F\' isica, Departamento de F\'{\i}sica, 
Universidade Federal Rural do Rio de Janeiro\\
BR 465-07, 23890-971, Serop\'edica, RJ, Brazil\\
${}^{b}$Departamento de F\'{\i}sica, ICE, Universidade Federal de Juiz de Fora,\\
36036-330, Juiz de Fora, MG, Brazil\\
\bigskip
\today\\
{\it Dedicated to the memory of Prof. Emerson da Silva Guerra}}


\begin{abstract}
\ni It is well known that a direct Lagrangian description of radiation damping is still missing.  In this paper we will use a specific approach of this problem which is the standard way to treat the radiation damping problem.
The objectives here are to construct: a $N=2$ supersymmetric extension for the model describing the radiation damping on the noncommutative plane with electric and magnetic interactions; a new dual equivalent action to the original one; the supercharge algebra and the total Hamiltonian for the system.
\end{abstract}

\maketitle

\pagestyle{myheadings}
\markright{Symmetry considerations on radiation damping}

\section{Introduction}
\label{intro}



An underlying feature of all charged particles is that an accelerated charged particle radiate electromagnetic energy.  During this process, the recoil momentum of the emitted photons is equivalent to a reaction force equivalent to the self-interaction of the particle with its own electromagnetic field which creates the radiation damping \cite{hjp,becker,Lorentz}.

The analysis of dissipative systems in quantum theory is of strong interest and relevance either because of fundamental reasons \cite{uz} or because of its practical applications \cite{cl,whb,walker,hk,Bula}.  
The explicit time dependence of the Lagrangian and Hamiltonian operators introduces a major difficulty in this study since the canonical commutation relations are not preserved by time evolution.  
Different approaches have been used in order to apply the canonical quantization scheme to dissipative systems \cite{mt,bc,fv,cl2,cl3,galley}.


Another way to handle with the problem of quantum dissipative systems is to double the target phase-space, so as we have to deal with an effective isolated system composed by the original system plus its time-reversed copy \cite{ft,bm,bm2}.  The new degrees of freedom thus introduced may be  represented by a single equivalent (collective) degree of freedom for the bath, which absorbs the energy dissipated by the system.  
In order to implement a canonical quantization formalism, we must first double the dimension of the target phase-space.  The objective of this doubling procedure is to comply with the canonical quantization scheme, which requires an effective isolated system.

To study the quantum dynamics of an accelerated charge, it is proper to use indirect representations since it loses the energy, the linear momentum and the angular momentum carried by the radiation field.  The effect of these losses to the motion of charge is known as radiation damping (RD) \cite{hjp}.

The reaction of a classical point charge to its own radiation was first discussed by Lorentz and Abraham more than a hundred years ago \cite{becker,Lorentz}.  There are two interesting aspects of the Abraham-Lorentz theory: the self-acceleration and pre-acceleration.  

Self-acceleration refers to the classical solutions where the charge is under acceleration even in the absence of an external field.  Pre-acceleration means that the charge begins to accelerate before the force begins to act.


So, a complete description of radiation damping is still missing.  In this way, in this paper we discussed some aspects of the RD framework concerning gauge symmetries, algebraic noncommutativity and supersymmetry, as well as the corresponding resulting physics, of course.  Notice that to talk about these issues in a RD system is very difficult because, since we have to deal with, in fact, two systems, the particle and the reservoir, both mathematical and physical features are not so trivial, as we will see.

We investigated the existence of a dual equivalent model to the RD one through the Noether method.  
We introduced a $N=2$ supersymmetric extension  for the RD model completing the $N=1$ supersymmetric version introduced in \cite{mendes,Barone}.  

We will describe a $N=2$ supersymmetric extension of the nonrelativistic $(2+1)$-dimensional pseudo-Euclidean space model describing the RD (represented by the equation (\ref{01}) below) on the noncommutative (NC) plane introducing an interaction to the free model by the $N=2$ superfield technique.  

However, it is important to notice that in fact there are two phase-spaces considered here.   The first one is where the radiation damping occurs and the second one is the double phase-space where details and relevance will be described in section II and in the references quoted there.   In this doubled phase-space, which we believe that the physics is not entirely understood, we performed the considerations described in this work, i. e., noncommutativity, duality and supersymmetry.  For example, concerning only noncommutativity, it can be shown easily that the original space is commutative whereas in this double phase-space we will show precisely in this paper, that it is NC.  We hope that our work can bring some light in the understanding of this extended space.

The organization of this paper is: in section 2 we will carry out a brief review of the mechanical model with a Chern-Simons term developed in \cite{Lukierski} and its Galilean-symmetric version, i.e., the LSZ model.  
We will obtain a dual equivalent model through the Noether dualization procedure in section 3.  In section 4  we will present a symplectic structure for the model in order to introduce the noncommutativity through the variables used in \cite{Plyushchay1,Plyushchay3}.  In section 5 we show the supersymmetric extension of the model, 
the supercharges and a supersymmetric version through the Hamiltonian formalism.  The conclusions and perspectives are described in the last section. 
 
\section{The model}

{\bf The LSZ model.} In \cite{Lukierski} the authors have introduced a nonrelativistic classical mechanics free particle model with a Chern-Simons-like term as
\be\label{01.1}
L_{LSZ}=\frac{1}{2}\,m\,{\dot x_i}^2 \,+\,\lambda\varepsilon_{ij} x_i \dot x_j, \;\;i,j=1,2,
\ee
where $\lambda$ has dimension of mass/time and the second term can be seen as a particular electromagnetic coupling of an electromagnetic potential.  This Lagrangian is neither invariant nor invariant under a total derivative under the Galilean boosts transformations.
A NC version of (\ref{01.1}) was studied in \cite{duval}.
To make (\ref{01.1}) quasi-invariant under $D=2$ Galilei symmetry the second term in (\ref{01.1}) was modified and we have that
\be \label{01.2}
L_{LSZ}=\frac{1}{2}\,m\,{\dot x_i}^2 \,-\,\kappa\varepsilon_{ij}\dot x_i \ddot x_j, \;\;i,j=1,2,
\ee
where $\kappa$ has dimensions of mass $\times$ time.  It can be shown \cite{ls} that this Lagrangian is quasi-invariant.  This model (\ref{01.2}) possess not a usual Galilei symmetry.  We can describe it by the exotic, two-fold centrally extended Galilei symmetry with non-commutating boosts. It was analyzed carefully in \cite{Lukierski} and later in \cite{duval2}.   
The authors in \cite{Lukierski} demonstrated that the model describes the superposition of a free motion in NC $D=2$ spaces.
A $N=2$ supersymmetric extension of (\ref{01.2}) was accomplished in \cite{Lukierski2} describing particles in the NC space with electric and magnetic interactions.  A supersymmetrization of (\ref{01.2}) was firstly obtained in \cite{lsz}. 
Both models above are depicted here to help the reader to understand the physical alternatives for actions like (\ref{01.1}) and (\ref{01.2}).  Other considerations can be found in \cite{ah}.



\bigskip

{\bf The radiation damping model.}  In \cite{Barone,Albert1} a new point of view concerning the study of RD \cite{hjp,hjp2,hjp3} was presented, introducing a Lagrangian formalism to the model in two dimensions given by
\be\label{01}
L_{RD}=\frac{1}{2}m\,g_{ij}\dot x_i \dot x_j -\frac{\gamma}{2}\varepsilon_{ij}\dot x_i \ddot x_j, \;\;i,j=1,2,
\ee
where $\varepsilon_{ij}$ is the Levi-Civita antisymmetric tensor, $g_{ij}$ is the metric for the pseudo-Euclidean plane \cite{bgpv} which is given by 
\be\label{02}
g^{ij}\,=\,g_{ij}=diag(1,-1)\,\,.
\ee
We are using the Einstein sum convention for repeated indices. The model (\ref{01}) was shown to have (1+1)-Galilean symmetry and the dynamical group structure associated with that system is {\it SU}(1,1) \cite{Albert1}.  {The supersymmetrization $N=1$ of (3) was studied in \cite{Albert2}. 

The Lagrangian (\ref{01}) describes, in this pseudo-Euclidean space, a dissipative system of a charge interacting with its own radiation, where the 2-system represents the reservoir or heat bath coupled to the 1-system \cite{mendes,Barone}.  It shows that the dissipative term, as a matter of fact, acts as a coupling term between the 1-system and the 2-system in this space.  In particular, we have a system including the charge and its time-reversed image, that globally behaves like a closed system described by equation (\ref{01}).

Note that the Lagrangian (\ref{01}) is similar to the one discussed in \cite{Lukierski} (action (\ref{01.2})), which is a special nonrelativistic limit of  the particle with torsion \cite{Plyushchay}. 
However, in this case we have a pseudo-Euclidean metric and the RD constant ($\gamma$) which act as a coupling constant of a Chern-Simons-like term. The RD constant $\gamma$ play the same mathematical role of the ``exotic" parameter $\kappa$ in (2) \cite{Lukierski,Lukierski2}.  However, there is an underlying physical difference between both $\gamma$ and $\kappa$.

It is important to reinforce that the difference between the results that will be obtained here and the ones in \cite{lsz} is that, besides the metric, the physical systems are different, where the RD constant $\gamma$ is not a simple coupling constant.  It depends on the physical properties of the charged particle, like the charge $e$ and mass $m$ which are related to the objects in its equations of motion which depicts an interaction between the charge and its own radiation field.  We will see that the RD constant introduced the NCY into the system,.


  
\section{Duality through Dualization}


The dualization technique \cite{iw,eu} is based on the traditional idea of a local lifting of a global symmetry and may be realized by an iterative embedding of Noether counter terms.  This technique was originally explored in the soldering formalism context \cite{abw,amw,bw} and was explored in \cite{iw2,ainrw,binrw} since it seems to be the most appropriate technique for non-Abelian generalization of the dual mapping concept.


Hidden symmetries may be revealed by a direct construction of a gauge invariant theory out of a non-invariant one \cite{nw,ai,monr,ainw}.  The advantage in having a gauge theory lies in the fact that the underlying gauge invariant theory allows us to establish a chain of equivalence among different models by choosing different gauge fixing conditions.  Clearly, the resulting embedded theory is dynamically equivalent to the original one \cite{ainrw}. This is the meaning of the ``duality" expression, namely dual equivalence.



As the first step, let us rewrite our RD model equation (\ref{01}),
\be
L_0^{RD}\,=\,{1\over 2}\,m\,g_{ij}\,\dot{x}_i\,\dot{x}_j\,-\,{\gamma\over 2}\,\varepsilon_{ij}\,\dot{x}_i\,\ddot{x}_j,\quad i,j=1,2\;\;,
\ee
hence the variation of this action is
\be \label{69}
\delta L_0^{RD}\,=\,J_1^i\,\dot{\eta}_i\,+\,J_2^i\ddot{\eta}_i
\ee
where $\delta x_i=\eta_i$ and the Noether currents are
\ba \label{7}
J_{1i}&=&m\,g_{ij}\,\dot{x}_j\,-\,{1\over 2}\,\gamma\,\varepsilon_{ij}\ddot{x}_j \\
J_{2i}&=&{1\over 2}\,\gamma\,\varepsilon_{ij}\dot{x}_j \;\;.
\ea
The second step in the iterative method \cite{iw,eu} is to construct the action with two new fields, i.e., the so-called auxiliary fields which will be eliminated through the equations of motion.  

Hence the new action is
\be \label{75}
L_1^{RD}\,=\,L_0^{RD}\,-\,D_1^i\,J_{1i}\,-\,D_2^i\,J_{2i}
\ee
where $D_1^i$ and $D_2^i$ are auxiliary fields.  


Let us establish another symmetry, namely,
\ba \label{76}
\delta D_1^i\,&=&\,\dot{\eta}_i \nonumber \\
\delta D_2^i\,&=&\,\ddot{\eta}_i \;\;.
\ea
With equations (\ref{69}) and (\ref{76}) we can carry out the variation of equation (\ref{75}), which results in,
\be \label{77}
\delta L_1^{RD}\,=\,-D^i_1\,(\delta J_{1i})\,-\,D^i_2\,(\delta J_{2i})\;\;.
\ee
Using the variations of the Noether currents, equations (\ref{7}) and (8), and that $\delta x_i = \eta_i$ we substitute these results in (\ref{77}) and we have that,
\be \label{80}
\delta L_1^{RD}\,=\,-\,m\,g_{ij}\,D_1^i\,\delta D_1^j\,+\,{1\over 2} \gamma\,\varepsilon_{ij}\,\delta(D_1^i\,D_2^j)\;\;.
\ee
To construct a gauge invariant action we have to use (\ref{80}) conveniently.  We can see directly that,
\be \label{90}
L_2^{RD}\,=\,L_1^{RD}\,+\,{1\over 2}\,m\,g_{ij}\,D_1^i\,D_1^j\,-\,{1\over 2} \gamma\,\varepsilon_{ij}\,D_1^i\,D_2^j
\ee
and accomplishing the variation of (\ref{90}), using (\ref{69}), (\ref{75}) and (\ref{80}) we can see clearly that it is gauge invariance so that $\delta L_2=0$ finishing the iterative chain.

From Eq. (\ref{90}) we can calculate the equations of motion for the auxiliary fields $D_1^1$, $D_1^2$, $D_2^1$ and $D_2^2$ and the results are
\ba \label{91}
D_1^1&=& -\,\frac{1}{\gamma}\, J_{22}  \nonumber \\
D_1^2&=&\frac{1}{\gamma}\, J_{21}   \nonumber \\
D_2^1&=&\frac{2}{\gamma}\,\Big[J_{12}\,+\,\frac{m}{\gamma}J_{21} \Big] \\
D_2^2&=&-\,\frac{2}{\gamma}\Big[ J_{11}\,+\,\frac{m}{\gamma}J_{22} \Big]\,\,, \nonumber 
\ea
where the Noether currents $J_{11},\,J_{12},\,J_{21}$ and $J_{22}$ are given by Eqs. (\ref{7}) and (8).  The next step is to substitute Eqs. (\ref{91}) into the Lagrangian in (\ref{90}).  The result is
\be \label{o}
L_2^{RD}\,=\, \frac 58 m (\dot{x}_1^2\,-\,\frac15 \dot{x}_2^2 ) \,+\,\frac 14 \gamma (\dot{x}_1 \ddot{x}_2\,-\,\dot{x}_2\ddot{x}_1 )
\ee
which can be rewritten as
\be \label{P}
L_2^{RD}\,=\, \frac 12 m (\dot{X}_1^2\,-\, \dot{X}_2^2 ) \,-\,\frac 12 \Gamma (\dot{X}_1 \ddot{X}_2\,-\,\dot{X}_2\ddot{X}_1 )
\ee
where
\ba \label{Q}
X_1&=&\frac{\sqrt{5}}{2}\,x_1 \nonumber\\
X_2&=&\frac{1}{2}\,x_2 \\
\Gamma&=&-\,\frac{2}{\sqrt{5}}\,\gamma \nonumber\\
\ea

\ni and $L_2^{RD}$ in Eq. (\ref{Q}) is invariant under $\delta X_i\,=\,\eta_i$ so we can say that $L_{RD}$ in Eq. (3) is self-dual..  Of course, $L_2^{RD}$ is the same as our original Lagrangian in Eq. (3).  We will talk about what this result means in a few moments.  Before that we would like to explore an connection involving $L_{LSZ}$ in Eq. (2), which can be explicitly written as
\be \label{R}
L_{LSZ}\,=\, \frac 12 m (\dot{x}_1^2\,+\, \dot{x}_2^2 ) \,-\,\kappa (\dot{x}_1 \ddot{x}_2\,-\,\dot{x}_2\ddot{x}_1 )
\ee
which is analogous to $L_{RD}$.  If we substitute 
\be \label{XXXX}
x_2 \rightarrow i\,x_2
\ee
\ni we have that
\be \label{S}
L_{LSZ}\,=\, \frac 12 m (\dot{x}_1^2\,-\, \dot{x}_2^2 ) \,-\,\frac 12 \kappa^{\prime} (\dot{x}_1 \ddot{x}_2\,-\,\dot{x}_2\ddot{x}_1 )
\ee
where $\kappa^{\prime}=\,-\,2 i \kappa$ in $L_{LSZ}$ above, which is equal to $L_{RD}$.  This result is a surprise because lead us to conclude directly that the Lagrangian $L_{LSZ}$ is dual to $L_{RD}$.  However, $L_{RD}$ is invariant under $\delta x_i\,=\,\eta_i$, but $L_{LSZ}$ is not.  Both Lagrangians are invariant under different gauge transformations.  This fact happens because, although in $L_{LSZ}$, $\lambda$ is simply a constant, in $L_{RD}$ $\gamma$ cannot be rewritten as in (\ref{R}) and (\ref{S}).   Since $\gamma$ and $\lambda$ are not connected, to find a Lagrangian dual to $L_{LSZ}$ we have to apply the Noether procedure from the beginning in Eq. (1).  This calculation is beyond the RD's  scope of this work.  These results show us that although, naively, we can think that both $L_{RD}$ and $L_{LSZ}$ are connected by the metric, i.e., $L_{RD}$ uses a pseudo-Euclidean metric, the Noether procedure confirms, once more, that they represent two completely different physical systems.  This fact is well known together with the fact that $\gamma$ in Eq. (3) is not a simple coupling constant.  It depends on the physical properties of the charged particle that is being analyzed.  To clarify, we can cite its charge (e) and mass (m).  Their relation with the terms is through the equation of motion, which depicts an interaction between the charge and its own radiation field.

Having said that, we can claim that what is new here is that $\gamma$ spoiled the connection between $L_{RD}$ and $L_{LSZ}$ through dualization, since one can think that this task (to obtain a dual action of one directly from the other's dual) would be an easy one.  Another consequence of Noether's procedure is that it also cannot relate $x_{2RD}$ and $x_{2LSZ}$ via Eq. (\ref{XXXX}) as we thought naively.

To sum up, we can say that although both $L_{RD}$ and $L_{LSZ}$ are mathematically analogous but the Noether approach showed precisely that they are physically different.

\section{Noncommutativity}

The study of NC theories has received a special attention through the last years thanks to the possibility that noncommutativity can explain 
the physics of the Early Universe.  
In other words the spacetime of the Early Universe can be a NC one.  It has been used in many areas of theoretical physics \cite{nekrasov}, cosmology \cite{nosso} and with Lorentz invariance \cite{amorim}. 

Introducing a Lagrangian multiplier which connects $\dot x$ to $z$, and substituting all differentiated $x$-variables in the Lagrangian 
(\ref{01}) by $z$-variables, one has a first-order Lagrangian
\be\label{03}
L^{(0)}\,=\,g_{ij} p_i (\dot x_j -z_j ) +\frac{m}{2}g_{ij} z_i z_j -\frac{\gamma}{2}\varepsilon_{ij}z_i \dot z_j\;\;.
\ee
The equations of motion can be written, using the symplectic structure \cite{Faddeev}, as
\be\label{04}
\omega_{ij}\dot \xi ^j =\frac{\partial H(\xi)}{\partial \xi^{i}}
\ee
where the symplectic two form is
\be\label{05}
(\omega)=
\begin{pmatrix}
{\bf0} & {\bf g} & {\bf0}\cr -{\bf g} & {\bf0} & {\bf0} \cr {\bf0} & {\bf0} & -\gamma \varepsilon 
\end{pmatrix}
\ee
with
\be\label{06}
\varepsilon=
\begin{pmatrix}
0 & 1 \cr -1 & 0 
\end{pmatrix},
\ee
where $g$ was given in Eq. (\ref{02}) and {\bf 0} denotes the $2\times\,2$ null matrix. $H(\xi^l)$ is the Hamiltonian and $\xi^i$ are the symplectic variables.

Using the variables introduced in \cite{Plyushchay1,Plyushchay3} modified as
\ba\label{07}
{\cal Q}_i &=&\gamma\,g_{ij}(mz_j \,-\,p_j )\;,  \nonumber \\
X_i&=&x_i \,+\,\varepsilon_{ij}{\cal Q}_{j}\;, \nonumber \\
P_i&=&p_i\,\,,
\ea
we can write that our Lagrangian can be separated into two disconnected terms in order to describe the ``external" and ``internal" degrees of freedom.  So, 
\be\label{08}
L^{(0)} =L^{(0)}_{\rm ext} + L^{(0)}_{\rm int}
\ee
where
\be\label{09}
L^{(0)}_{\rm ext}\,=\,g_{ij}\,P_i \dot X_j + {\gamma\over2}\varepsilon_{ij}P_i \dot P_j -{1\over 2m}g_{ij}P_i P_j ,
\ee
and
\be\label{10}
L^{(0)}_{\rm int}={1\over{2\gamma}}\varepsilon_{ij}{\cal Q}_i \dot{\cal Q}_j +{1\over{2m\gamma^2}}g_{ij}{\cal Q}_i {\cal Q}_j .
\ee

The internal coordinates, $\vec{{\cal Q}}$, and the external ones, $\vec{X}$, are decoupled \cite{Plyushchay1} and we can see that they do not commute,
with the following nonvanishing Poisson brackets,
\ba\label{11}
\{X_i ,X_j \} &=&\gamma \varepsilon_{ij},\;\;\; \{X_i ,P_j \}\,=\,g_{ij},\nonumber\\
\{{\cal Q}_i ,{\cal Q}_j \}&=& \gamma\varepsilon_{ij}\,\,.
\ea

We can see in (\ref{08}) that our Lagrangian can be written as two separated and disconnected parts describing the ``external'' and ``internal'' degrees of freedom in a NC phase space, parameterized by the variables $(X_i ,P_i)$ (external structure) and ${\cal Q}_i$(internal structure) 
\cite{Plyushchay1,Plyushchay3}. 

Now we introduce an interaction term to the ``external'' sector, equation (\ref{09}), which do not modify the internal sector, represented by a potential energy term $U(X)$ involving NC variables, as follows
\be\label{11.0}
L_{\rm ext}\,=\,g_{ij}\,P_i \dot X_j + {\gamma\over2}\varepsilon_{ij}P_i \dot P_j -{1\over 2m}g_{ij}P_i P_j -U(X)\,\,.
\ee
This leads to a deformation of the constraint algebra, since the constraint now involves a derivative of the potential \cite{Albert1}.

To end this section, notice the $L^{(0)}$ has the double of the phase-space dimension.  The ``external" and ``internal" dynamics should be interpreted in terms of underlying one-dimensional dissipative dynamics.  The same observation can be made for the results below.

Notice that the Lagrangian in Eqs. (27), (28) and (30) are formed by the objects that shows a NC algebra described in Eq. (29).  The standard NC procedure is to recover the commutative algebra in (29) using the so-called Bopp shift.
\be \label{AXXXXX}
x_i \,=\, \hat{x}_i \,+\, \frac 12 \gamma \epsilon_{ij} p_j
\ee
where the hat defines a NC variable.  Using (\ref{AXXXXX}) in (29) we will have that $\{x_i , x_j \} = 0$.  The same can be made with ${\cal Q}_i$ so that
\be \label{BXXXXX}
{\cal Q}_i\,=\,\hat{{\cal Q}}_i\,+\,\frac 12 \gamma \epsilon_{ij} {\cal P}_j
\ee
where $\hat{p}_i\,=\,p_i$ and ${\cal P}_i\,=\,{\hat{\cal P}}_i$ and consequently $\{p_i , p_j \}=\{{\cal P}_i , {\cal P}_j\}=0$.   Substituting (\ref{AXXXXX}) and (\ref{BXXXXX}) in the Lagrangians (27), (28) and (30) results in a Lagrangian defined in NC phase-space.    But, to go further in this analysis is an ongoing research.

\section{The supersymmetric model in $N=2$}

To obtain the supersymmetric extension of the model described by the Lagrangian (\ref{11.0}), for each commuting space coordinate, representing the system degrees of freedom, we will associate one anti-commuting variable, which are the well known Grassmannian variables. We are considering only the $N=2$ BUSY for a non-relativistic particle, which is described by the introduction of two real Grassmannian variables $\Theta$ and $\bar\Theta$ (the Hermitian conjugate of $\Theta$) in the configuration space, but all the dynamics are represented by the time $t$ \cite{Galvão,Junker}.

Let us carry out the Taylor expansion for the real scalar supercoordinate as
\ba
\label{11.1}
&X_i& \rightarrow {\cal X}_i(t,\Theta,\bar\Theta)= \nonumber \\
&=&X_i(t) + i\psi_i(t)\Theta + i\bar\Theta \bar{\psi}_i(t) +\bar{\Theta}\Theta F_i(t)
\ea
and their canonical supermomenta
\ba\label{11.2}
&P_i(t)& \rightarrow {\cal P}_i(t,\Theta,\bar\Theta)= \nonumber \\
&=&i\eta_i(t) - i\Theta\left(P_i(t)+if_i(t)\right)-\bar{\Theta}\Theta\dot\eta_i(t),
\ea
which under the infinitesimal supersymmetry transformation law
\ba\label{11.2.0}
\delta t=i\bar{\epsilon}\theta + i\bar{\epsilon}\Theta, \;\;\; \delta\Theta=\epsilon \;\;\;{\rm and}\;\;\; \delta\bar{\Theta}=\bar{\epsilon},
\ea
where $\epsilon$ is a complex Grassmannian parameter, we can write that 
\ba\label{11.2.1}
\delta {\cal X}_i &=&(\epsilon\bar{Q} + \bar{\epsilon}Q){\cal X}_i \\
\mbox{and} \qquad \delta{\cal P}_i &=& (\epsilon\bar{Q} + \bar{\epsilon}Q){\cal P}_i\,\,,
\ea
where both $Q$ and $\bar{Q}$ are the two SUSY generators
\be\label{11.2.3}
Q=\frac{\partial}{\partial \bar{\Theta}}+i\Theta\frac{\partial}{\partial t},\;\;\;\;\; \bar{Q}=-\frac{\partial}{\partial {\Theta}}-i\bar{\Theta}\frac{\partial}{\partial t}.
\ee

In terms of $(X_i(t), P_i(t), F_i,f_i)$, the bosonic (even) components and $(\psi_i(t),\bar{\psi}_i(t),\eta_i(t))$, the fermionic (odd) components, we obtain the following supersymmetric transformations,
\ba\label{11.2.4}
\delta X_i \,=\, i(\bar{\epsilon}\bar{\psi}_i +\epsilon\psi_i ) &;&\quad 
\delta \psi_i \,=\, -\bar{\epsilon}(\dot{X}_i -iF_i )\nonumber\\
\delta \bar{\psi}_i \,=\, -\epsilon (\dot{X}_i +iF_i ) &;&\quad  
\delta F_i \,=\, \epsilon\dot{\psi}_i - \bar{\epsilon}\dot{\bar{\psi}}_i \,,
\ea
and
\be\label{11.2.5}
\delta\eta_i = \epsilon(P_i +if_i) ;\quad
\delta P_i = 0 ;\quad
\delta f_i = 2\bar{\epsilon}\dot{\eta}_i\;\;. 
\ee 
%

The super-Lagrangian for the super point particle with $N=2$, invariant under the transformations (\ref{11.2.4}) and (\ref{11.2.5}),  can be written as the following integral (we use for simplicity that $m=1$)
\ba\label{11.3}
\bar{L}_{\rm ext}&=&\frac{1}{2}\int{\rm d}\Theta {\rm d}{\bar\Theta}\left[ \frac{}{}g_{ij}\(\bar{D}{\cal X}_i \bar{\cal P}_j+{\cal P}_jD{\cal X}_i\)\right. \nonumber\\ 
&+&\left. \frac{\gamma}{2}\varepsilon_{ij}\({\cal P}_i\,{\dot{\bar{\cal P}}}_j \,+\,{\dot{\cal P}}_j \,{\bar{\cal P}}_i \) -\frac{1}{2}g_{ij}\({\cal P}_i\bar{\cal P}_j +{\cal P}_j\bar{\cal P}_i \)\right]\nonumber\\
&-& \int{\rm d}\,\Theta {\rm d}\,{\bar\Theta}U[{\cal X}(t,\Theta,\bar\Theta )]
\ea
where $D$ is the covariant derivative $(D=\partial_{\Theta}-i\bar\Theta\partial_t)$ and $\bar D$ is its Hermitian conjugate. The $U[{\cal X}]$ is a polynomial function of the supercoordinate

Expanding  the superpotential $U[{\cal X}]$ in Taylor series and maintaining $\Theta\bar\Theta$ (because only these terms survive after  integrations on Grassmannian  variables $\Theta$ and $\bar\Theta$), we have that
\ba\label{11.4}
U[{\cal X}]&=&{\cal X}_i \frac{\partial U[X(t)]}{\partial X_i} + \frac{{\cal X}_i{\cal X}_j^*}{2}\frac{\partial^2 U[X(t)]}{\partial X_i \partial X_j}+ ...\\
&=&F_i \bar\Theta \Theta \partial_i U[ X(t)]+\bar\Theta \Theta \psi_i \bar{\psi}_j \partial_i \partial_j U[ X (t)]+...\nonumber
\ea
where the derivatives $\partial_i =\frac{\partial}{\partial X_i}$ are such that $\Theta=0=\bar\Theta$, which are functions only of the $X(t)$ even coordinate. Substituting equation (\ref{11.4}) in equation (\ref{11.3}), we  obtain after integrations 
\ba\label{11.5}
\bar{L}_{\rm ext}&=& L^{(0)}_{\rm ext} -\frac{1}{2}g_{ij}f_i f_j-g_{ij}F_i f_j +\frac{\gamma}{2}\varepsilon_{ij}f_i\dot{f}_j \nonumber\\ &-&big_{ij}\(\bar{\psi}_i \dot{\bar \eta}_j  -\dot{\eta}_j\psi_i \) -big_{ij}\dot{\eta}_i \bar{\eta}_j +i\gamma \varepsilon_{ij}\dot{\eta}_i\dot{\bar\eta}_j \nonumber\\&-&F_i \partial_i U[X(t)]-\psi_i \bar{\psi}_j \partial_i \partial_j U[X (t)],
\ea
which is the complete Lagrangian for $N=2$. 

The bosonic component $F_i$ is not a  dynamic variable. In this case, using the Euler-Lagrange equations for the auxiliary variables $f_i$ and $F_i$, we obtain
\ba
f_i(t)&=& g_{ij}\partial_j U[X(t)],\label{11.6}\\
F_i(t)&=& f_i +\gamma g_{ail}\varepsilon_{j}\dot f_j \nonumber\\
&=&g_{ij} \partial_j U[X(t)]-\gamma\varepsilon_{ij} \partial_j\partial_k U[X] \dot X_k (t),\label{11.7}
\ea
where we have to eliminate the variable $f_i$ as well as its derivative in $F_i$.  Now, substituting (\ref{11.6}) and (\ref{11.7}) in  (\ref{11.5}) the auxiliary variables can be completely eliminated, hence 
\ba\label{11.8}
\bar{L}_{(N=2)\rm{ext}}&=&L_{\rm{ext}}^{(0)} -\frac{1}{2}g_{ij}\partial_i U\partial_j U+\frac{\gamma}{2}\varepsilon_{ij}\partial_i U\partial_j\partial_k U\dot X_k \nonumber \\&-&big_{ij}\(\bar{\psi}_i \dot{\bar \eta}_j  -\dot{\eta}_j\psi_i \) -big_{ij}\dot{\eta}_i \bar{\eta}_j + i\gamma \varepsilon_{ij}\dot{\eta}_i\dot{\bar\eta}_j\nonumber\\ &-&\psi_i\bar{\psi}_j \partial_i\partial_j U\,\,.
\ea

Note that, as in \cite{Lukierski2}, we can rewrite equation (\ref{11.8})
 as 
\ba\label{11.9}
\bar{L}_{(N=2)\rm{ext}}&=&{L}_{\rm{ext}}^{(0)} +A_k(X,t)\dot X_k +A_0(X,t)+ \nonumber \\&-&big_{ij}\(\bar{\psi}_i \dot{\bar \eta}_j  -\dot{\eta}_j\psi_i \) -big_{ij}\dot{\eta}_i \bar{\eta}_j + i\gamma \varepsilon_{ij}\dot{\eta}_i\dot{\bar\eta}_j\nonumber\\ &-&\psi_i\bar{\psi}_j\partial_i\partial_j U,
\ea
which is invariant under standard gauge transformations $A_{\mu}\rightarrow A_{\mu}^{\prime}=A_{\mu} +\partial_{\mu}\Lambda$, where
\be\label{11.10}
A_0( X,t) =-\frac{1}{2}g_{ij}\partial_i U\partial_j U
\ee
and 
\be\label{11.12}
A_k(X,t)=\frac{\gamma}{2}\varepsilon_{ij}\partial_i U\partial_j\partial_k U\,\,
\ee
were both identified in \cite{Lukierski2} with the scalar potential $A_0$ (that in this case have a pseudo-Euclidean metric) and the vector potential $A_k$.  
Notice that both potentials above are not independent.
The vector potential introduce a magnetic field $B=\varepsilon_{ij}\partial_i A_j$ given by
\be\label{11.13}
B( X)=\frac{\gamma}{2}\varepsilon_{AK}\varepsilon_{j}\left(\partial_i\partial_l U\right)\left(\partial_j\partial_k U\right)
\ee
where we can see that the noncommutativity introduced by the parameter $\gamma$ generates a constant magnetic field \cite{Lukierski2} and an electric field given by $E_i\,=\,\p_i\,A_0$ which can be written as
\be\label{11.13.1}
E_i\,(X)\,=\,-\,g_{j}\,\p_i\,\p_l\,U\,\p_j\,U\,\,.
\ee

The Euler-Lagrange equations, in this case, are
\begin{subequations}
\label{11.14}
\ba
m^* \dot X_i \,=\, P_i \,&-&\,me\gamma\varepsilon_{ij}E_j \nonumber \\
\,&+&\, m\gamma\varepsilon_{ij}\psi_l\bar{\psi}_k\partial_l\partial_k\partial_j U, \label{11.14a}\\
\dot P_i \,=\, e\,g_{ij}\varepsilon_{Al}\dot X_{l}B &+& beg_{ij}E_j \nonumber \\
&-& g_{ij}\psi_l\bar{\psi}_k\partial_l\partial_k\partial_j U,\label{11.14b}
\ea
\end{subequations}
where $E_i$ and $B$ are the electric and magnetic field, respectively, and $m^{*} =m(1-e\gamma B)$ is an effective mass. However, this way of introducing electromagnetic interaction modifies the symplectic structure of the system which determines the NC phase-space geometry, for the bosonic sector, equation (\ref{11}), we have
\ba\label{11.15}
\{X_i,X_j \}&=&\frac{m}{m^*}\gamma\varepsilon_{ij},\;\; \{X_i ,P_j \}=\frac{m}{m^*}g_{ij},\nonumber\\
\{P_i ,P_j \}&=&\frac{m}{m^*}b\varepsilon_{ij},
\ea
which imply an analysis of the value $e\gamma B \neq 1$ in order to avoid a singularity \cite{Plyushchay2,duval}.  Notice that the algebra in (\ref{11.15}) is different from the one in (29) where the momenta commute.

Concerning the fermionic sector, the Euler-Lagrange equations are
\ba\label{11.16}
i\gamma\varepsilon_{ij}{\ddot{\bar\eta}}_j +big_{ij}{\dot{\bar\eta}}_j -i\dot\psi_i &=&0,\nonumber\\
i\gamma\varepsilon_{ij}{\ddot{\eta}}_j +big_{ij}{\dot\eta}_j -i\dot{\bar\psi_i} &=&0,
\ea
for the fermionic variables $(\eta,\bar\eta)$. For the fermionic variables  $(\psi_i ,\bar{\psi}_i )$ the Euler-Lagrange equations are
\ba\label{11.17}
i\dot\eta_i + g_{AK}\bar\psi_j \partial_k\partial_j U &=& 0\nonumber\\
i\dot{\bar\eta}_i - g_{AK}\psi_j \partial_j \partial_k U &=&0\,\,.
\ea
where the fermionic variables $(\psi_i, \bar{\psi}_i )$ do not have dynamics.

So, analogously to \cite{Lukierski2} we have here that the noncommutativity originates electric and magnetic fields.  In the case of RD, studied here, in the NC hyperbolic phase-space, the movement of the charged particle has an extra electromagnetic energy that did not appear in an $N=1$ SUSY analysis \cite{Albert2}.  This result agree with the fact that noncommutativity does not change the physics of the system.  However, we understand that this electromagnetic energy is an extra one due to the NC feature of the phase-space.  This result is also different, as it should be expected, from the one obtained in \cite{Lukierski2} where only a magnetic interaction appear.

\subsection{The harmonic oscillator solutions}

In order to obtain an interesting solution of equations (\ref{11.14}) let us consider a particular form for the superpotential like,
\be\label{446}
U(X)\,=\,\frac{\omega}{2}\,g_{ij}\,X_i\,X_j\,\,,
\ee
which has clearly an harmonic-like form.

It is easy to see that in both equations (\ref{11.14a}) and (\ref{11.14b}) the last term with three derivatives disappear and so we have two new equations with the fermionic and bosonic sector separated so that,
\begin{subequations}
\label{447}
\ba
& &m^*\,\dot{X}_i \,=\, P_i\,-\,e\,\gamma\,\epsilon_{ij}\,E_j \label{447a}\\
& &\!\!\!\!\!\!\!\!\!\!\!\!\!\!\!\!\!\!\!\!\!\!\!\!\!\!\!\!\!\!\!\!\!\!\!\!\!\!\!\!\!\!\!\!\!\!\!\!\!\!\!\!\!\!\!\!\!\!\!\!\!\!\!\!\!\!\!\!\!\!\!\!\!\!\!\!\!\!\!\!\!\!\!\mbox{and} \nonumber \\
& &\dot{P}_i \,=\, e\,g_{Al}\,\epsilon_{ij}\,\dot{X}_l\,B\,+\,e\,g_{ij}\,E_j \label{447b}\,\,.
\ea
\end{subequations}

\ni Computing a second time derivative of equation (\ref{447a}) we have
\be\label{448}
m^*\,\ddot{X}_i\,=\,\dot{P}_i\,-\,e\,\gamma\,\epsilon_{ij}\,\dot{E}_j\,\,.
\ee

\ni Substituting (\ref{447b}) into (\ref{448}) we have that
\be\label{449}
m^*\,\ddot{X}_i\,=\,  e\,g_{ij}\,\epsilon_{Al}\,\dot{X}_l\,B\,+\,e\,g_{ij}\,E_j \,-\,e\,\gamma\,\epsilon_{ij}\,\dot{E}_j\,\,,
\ee

\ni which will disclose a very well known result in a moment.

Back in (\ref{11.10}) and (\ref{11.12}) but now using (\ref{446}) we can write that,
\begin{subequations}
\label{450}
\ba
A_0\,(X,t) &=& -\,\frac{\omega^2}{2}\,g_{ij}\,X_i\,X_j \,\,,\label{450a}\\
A_k\,(X,t) &=& \frac{\gamma}{2}\,\omega^2\,\epsilon_{j}\,X_j\label{450b}
\ea
\end{subequations}

\ni respectively.  Substituting these equations in (\ref{11.13}) and (\ref{11.13.1}) we have that
\be\label{551}
B\,=\,\gamma\,\omega^2
\ee
and
\be\label{552}
E_i\,=\,-\,\omega^2\,g_{ij}\,X_j\,\,,
\ee

\ni and finally, substituting these both equations in (\ref{449}) it is easy to show that
\be\label{553}
\ddot{X}_i\,-\,\frac{\gamma\,e\,\omega^2}{1-\gamma^2\,\omega^2\,e}\,(\,g_{ij}\,\epsilon_{AK}\,+\,g_{AK}\,\epsilon_{ij}\,)\,\dot{X}_k 
\,+\,\frac{e\,\omega^2}{1-\gamma^2\,\omega^2\,e}\,X_i\,=\,0\,\,,
\ee

\ni which is the equation of a damped harmonic oscillator and we see clearly that the second-term of (\ref{553}) represents a dissipative force proportional to the velocity and in the last term of (\ref{553}), we have that
$$\omega_0^2\,=\,\frac{e\,\omega^2}{1-\gamma^2\,\omega^2\,e}\,\,$$

\ni represents the natural frequency of this oscillator $\omega_0$.  Notice that the RD constant is responsible for the dissipative force and  affects the frequency also.    The instantaneous rate of energy of the oscillator in equation (\ref{553}) can be written as
\be\label{554}
\frac{dE}{at}\,=\,m^*\,\frac{\gamma\,e\,\omega^2}{1-\gamma^2\,\omega^2\,e}\,\dot{X}_i\,\,,
\ee
so that the RD constant also affects the energy rate.  However, we note that equation (\ref{553}) has a general metric so that this equation is a general case.
Using the metric for the pseudo-Euclidean plane given in (4) we see that the second term in (\ref{553}) disappear, and we have that
$$\ddot{X}_i\,+\,\frac{e\,\omega^2}{1-\gamma^2\,\omega^2\,e}\,X_i\,=\,0$$
$$\Longrightarrow \ddot{X}_i\,+\,\omega_0^2\,X_i\,=\,0\,\,,$$

\ni which is the equation for the standard harmonic oscillator which has the standard solutions.

From (\ref{553}) we can see that, since there is not a term which has three derivatives of $X$, one can conclude that in the NC space, the non-physical solutions, namely the pre-acceleration solutions (for $\dddot{X}$), do not exist.

\subsection{The supercharge algebra}

Now, from the supersymmetric transformations, equations (\ref{11.2.4}) and (\ref{11.2.5}), and the Lagrangian (\ref{11.9}), we can compute the supercharge, through the Noether's theorem. The results for the charge operator are given by
\be\label{11.17.1}
Q=big_{ij}(P_i -i_i )\psi_j \;\;\; {\rm and} \;\;\; \bar{Q}=big_{ij}(P_i + i_i)\bar{\psi}_j \;\;,
\ee
where $W_i (X) =\partial_i U(X)$.

The supercharge algebra is
\be\label{11.7.2}
\{Q,\bar{Q}\}=\{\bar{Q},Q\} =0,
\ee
and
\be\label{11.7.3}
\{Q,\bar{Q}\} =-2i H .
\ee
Further, we easily carried out a canonical calculation of the Hamiltonian and we find that
\be\label{11.7.4}
H=H_{b} + H_{f},
\ee
where the bosonic Hamiltonian $H_{b}$ is
\be\label{11.7.5}
H_{b}=\frac{1}{2}g_{ij} \(P_i P_j + W_i W_j \)\; ,
\ee
and the fermionic part $H_f$ is
\be\label{11.7.6}
H_f = \frac{m}{m^{*}}\( ieB(X)\varepsilon_{ij} \bar{\psi}_i \psi_j +g_{ik}\partial_j W_k (X) \bar{\psi}_i \psi_j \).
\ee
Note that the second term in $H_b$ is proportional to the scalar potential, equation (\ref{11.10}), i. e., there is a potential energy term in $H_b$.  We can say that the origin of this term is related to the electric field.

There is an alternative way to introduce the minimal electromagnetic interaction.  It can be accomplished through the transformation $P_i \rightarrow {\cal P}_i =P_i + eA_i (X_i ,t)$ in the Hamiltonian, that preserve the symplectic structure of equation (\ref{11}). In \cite{Lukierski2} this transformation has been considered and it leads to the same expression for the magnetic field equation (\ref{11.13}).

\section{Remarks and conclusions}

A fundamental property of all charged particles is that the electromagnetic energy is radiated whenever they are accelerated.  The recoil momentum of the photons emitted during this process is equivalent to a reaction force corresponding to self-interaction of the particle with its own  electromagnetic field, which originates the RD.

 

Here the supersymmetric model was split into ``external" and ``internal" degrees of freedom of the supersymmetric model in terms of new variables, where the RD constant introduced noncommutativity in the coordinate sector.
We presented a way to introduce an electromagnetic coupling.  

We performed the supersymmetric $N=2$ extension of the RD model and realized that the noncommutativity introduced by the parameter generates a constant magnetic field.  With this result, together with the electric field we obtained a general expression for the damped harmonic oscillator which results in the standard harmonic oscillator in our pseudo-Euclidean space.  We saw that in the NC space, the non-physical solutions, namely the pre-acceleration solutions disappear.  After that we compute the supercharges algebra and the total Hamiltonian of the system, separated in bosonic and fermionic parts.

Also in this work, we used an alternative way to construct a dual equivalent action to the RD one, a dualization procedure.  We showed that the RD action is self-dual and also that, despite LSZ can be transformed in the RD action, both have different symmetries.  The dualization procedure showed precisely that although both actions are mathematically equivalent, they are physically different thanks to its coupling constant.  The RD one depends on each problem while for the LSZ action, it is simply a constant parameter.   Although it can sound like an obvious thing, but it is not.

With this new features revealed here we hope that this work has improved the fathoming of this extended space, which we believe it is not closed in the current literature.  

A perspective for future analysis is to study some typical problems of dissipative systems, like self-acceleration and pre-acceleration, for example.   To accomplish this, in our $N=2$ supersymmetric case, we have to begin analyzing the Euler-Lagrange equations (\ref{11.14}), (\ref{11.16}) and (\ref{11.17}).

\section{Acknowledgments}

The authors would like to thank CNPq (Conselho Nacional de Desenvolvimento Cient\' ifico e Tecnol\'ogico) and FAPEMIG (Funda\c{c}\~ao de Amparo \`a Pesquisa do 
Estado de Minas Gerais), Brazilian scientific support agencies, for partial financial support.

\end{document}